\PassOptionsToPackage{unicode}{hyperref}
\PassOptionsToPackage{hyphens}{url}
\documentclass[a4paper]{article}
\usepackage{amsmath,amssymb}
\usepackage{iftex}
\ifPDFTeX
  \usepackage[T1]{fontenc}
  \usepackage[utf8]{inputenc}
  \usepackage{textcomp} 
\else 
  \usepackage{unicode-math} 
  \defaultfontfeatures{Scale=MatchLowercase}
  \defaultfontfeatures[\rmfamily]{Ligatures=TeX,Scale=1}
\fi
\usepackage{lmodern}
\ifPDFTeX\else
\fi
\IfFileExists{upquote.sty}{\usepackage{upquote}}{}
\IfFileExists{microtype.sty}{
  \usepackage[]{microtype}
  \UseMicrotypeSet[protrusion]{basicmath} 
}{}
\makeatletter
\@ifundefined{KOMAClassName}{
  \IfFileExists{parskip.sty}{%
    \usepackage{parskip}
  }{
    \setlength{\parindent}{0pt}
    \setlength{\parskip}{6pt plus 2pt minus 1pt}}
}{
  \KOMAoptions{parskip=half}}
\makeatother
\usepackage{xcolor}
\usepackage{longtable,booktabs,array}
\usepackage{calc} 
\usepackage{etoolbox}
\makeatletter
\patchcmd\longtable{\par}{\if@noskipsec\mbox{}\fi\par}{}{}
\makeatother
\IfFileExists{footnotehyper.sty}{\usepackage{footnotehyper}}{\usepackage{footnote}}
\makesavenoteenv{longtable}
\providecommand{\tightlist}{%
  \setlength{\itemsep}{0pt}\setlength{\parskip}{0pt}}
\ifLuaTeX
  \usepackage{selnolig}  
\fi
\IfFileExists{bookmark.sty}{\usepackage{bookmark}}{\usepackage{hyperref}}
\IfFileExists{xurl.sty}{\usepackage{xurl}}{} 
\hypersetup{
  hidelinks,
  pdfcreator={Raul Brajczewski Barbosa}}

\author{Raul Brajczewski Barbosa\\
University of Coimbra, Portugal}
\date{November 24, 2023}
\title{\LARGE Petit programming language and compiler}

\newcommand{\removelabel}[1]{}

\begin{document}

\maketitle

Petit is an educational programming language for learning compilers. Students embark on the journey of learning compilers through a series of six tutorials\footnote{The complete source code is available at https://github.com/rbbarbosa/Petit}, progressing from topics like lexical analysis and syntactic analysis to semantic analysis and code generation. The initial tutorials in this series cover the practical applications of the lex and yacc tools, while the concluding tutorial focuses on generating LLVM intermediate representation.

In the Petit language, programs are comprised of functions which, in turn, are constructed with expressions. Here's a comprehensive example which covers most of the features:

\begin{verbatim}
factorial(integer n) = if n then n * factorial(n-1) else 1

main(integer i) = write(factorial(read(0)))
\end{verbatim}

\setcounter{tocdepth}{1}
\tableofcontents

\newpage
\hypertarget{compilers-tutorial-i-lexical-analysis}{%
\section{Compilers tutorial I: Lexical
analysis}\removelabel{compilers-tutorial-i-lexical-analysis}}

\emph{Lex} is a powerful tool commonly used in text processing and
compiler construction. It is designed to generate lexical analysers,
also known as lexers or tokenisers. By specifying patterns using regular
expressions, \emph{lex} allows us to instruct the analysers to recognise
and process specific tokens in the input text.

A bit of theory: \emph{Lex} takes the user-specified regular expressions
as input and applies an algorithm, such as Thompson's construction
algorithm, to convert them into equivalent NFAs (non-deterministic
finite automata). These NFAs are then converted into DFAs (deterministic
finite automata) using the subset construction algorithm. The DFA
generated by \emph{lex} is represented internally as a transition table,
which captures the state transitions based on input characters, enabling
efficient and deterministic token recognition during lexical analysis.

\hypertarget{using-lex}{%
\subsection{\texorpdfstring{Using
\emph{lex}}{Using lex}}\removelabel{using-lex}}

To use \emph{lex}, we specify patterns using regular expressions, along
with corresponding actions. \emph{Lex} then transforms these rules into
a C program that functions as the lexical analyser. When the lexical
analyser is executed, it scans the input text, recognises the patterns
specified in the rules and triggers the corresponding actions which are
user-written code fragments.

A \emph{lex} source file has three sections separated with the
\texttt{\%\%} delimiter:

\begin{verbatim}
...definitions...
%% 
...rules...
%% 
...subroutines...
\end{verbatim}

The \emph{rules} section contains our lexical specification: regular
expressions matching the patterns we are interested in, paired with
fragments of C code. A simple specification is:

\begin{verbatim}
%%
[0-9]+                      { printf("NATURAL\n"); }
\end{verbatim}

This \emph{lex} specification matches natural numbers, found in the
input text, and the corresponding action is to print the word
\texttt{NATURAL} each time. Any other unspecified patterns are directly
copied to the output without modification.

We can add a rule for decimal numerals:

\begin{verbatim}
%%
[0-9]+                      { printf("NATURAL\n"); }
[0-9]*"."[0-9]+             { printf("DECIMAL\n"); }
\end{verbatim}

If we ran this analyser, it would replace all naturals with the word
\texttt{NATURAL} and all decimals with the word \texttt{DECIMAL},
leaving any other characters unchanged.

In the \emph{definitions} section that comes before the rules, we can
place abbreviations to avoid repetitions and make specifications easier
to read. For example, we can use \texttt{\{digit\}} instead of
\texttt{{[}0-9{]}} by placing \texttt{digit} in the definitions section.
In the \emph{subroutines} section that comes after the rules, we write
any C functions we need, typically including functions \texttt{main()}
and \texttt{yywrap()}. Therefore, our first complete example is as
follows:

\begin{verbatim}
digit   [0-9]
%%
{digit}+                { printf("NATURAL\n"); }
{digit}*"."{digit}+     { printf("DECIMAL\n"); }
%%
extern int yylex();
int main() {
    yylex();    /* run the lexical analysis automaton */
    return 0;
}
int yywrap() {  /* called on EOF, return 1 to terminate */
    return 1;
}
\end{verbatim}

\hypertarget{generating-and-running-the-lexical-analyser}{%
\subsection{Generating and running the lexical
analyser}\removelabel{generating-and-running-the-lexical-analyser}}

Having the above specification in a file named \texttt{lexer.l}, we
obtain the C code for the lexical analyser by entering:

\begin{verbatim}
$ lex lexer.l
\end{verbatim}

The generated source code is written to a file called \texttt{lex.yy.c}
by default. We simply compile it using a C compiler:

\begin{verbatim}
$ cc lex.yy.c -o lexer
\end{verbatim}

The resulting executable file \texttt{lexer} reads from \texttt{stdin}
and writes to \texttt{stdout}. We can then run the analyser:

\begin{verbatim}
$ ./lexer
\end{verbatim}

Try it with integers, decimals and other tokens.

A bit of theory: The transition table which represents the lexical
analysis DFA is placed in the \texttt{lex.yy.c} file (around source line
\textasciitilde400). If it weren't for \emph{lex} we would have to
manually create those tables.

\hypertarget{regular-expressions}{%
\subsection{Regular expressions}\removelabel{regular-expressions}}

When it comes to regular expressions, different tools may have slight
variations in notation. For example, \emph{lex} uses a notation where
special characters are preceded by a backslash. The table below
summarises the main notations used by \emph{lex}.

\begin{longtable}[]{@{}
  >{\raggedright\arraybackslash}p{(\columnwidth - 4\tabcolsep) * \real{0.1618}}
  >{\raggedright\arraybackslash}p{(\columnwidth - 4\tabcolsep) * \real{0.5294}}
  >{\raggedright\arraybackslash}p{(\columnwidth - 4\tabcolsep) * \real{0.3088}}@{}}
\toprule\noalign{}
\begin{minipage}[b]{\linewidth}\raggedright
Expression
\end{minipage} & \begin{minipage}[b]{\linewidth}\raggedright
Description
\end{minipage} & \begin{minipage}[b]{\linewidth}\raggedright
Examples
\end{minipage} \\
\midrule\noalign{}
\endhead
\bottomrule\noalign{}
\endlastfoot
\texttt{x} & Character x & \texttt{a,\ 1} \\
\texttt{\textbackslash{}x} & x, if x is a lex operator &
\texttt{\textbackslash{}",\ \textbackslash{}\textbackslash{}} \\
\texttt{"xy"} & xy, even if x or y are lex operators &
\texttt{".",\ "*",\ "/*"} \\
\texttt{{[}x-z{]}} & Any character from x to z & \texttt{{[}0-9{]}},
\texttt{{[}a-z{]}} \\
\texttt{{[}xy{]}} & Either character x or y &
\texttt{{[}abc{]},\ {[}-+{]},\ {[}eE{]}} \\
\texttt{xx\textbar{}yy} & Either xx or yy &
\texttt{cat\textbar{}dog,\ if\textbar{}else} \\
\texttt{{[}\^{}x{]}} & Any character except x &
\texttt{{[}\^{}\textbackslash{}n{]},\ {[}\^{}a{]}} \\
\texttt{.} & Any character except newline & \\
\texttt{\^{}x} & x, at the start of a line & \\
\texttt{x\$} & x, at the end of a line & \\
\texttt{x*} & x, repeated 0 or more times & \texttt{{[}0-9{]}*,\ a*} \\
\texttt{x+} & x, repeated 1 or more times &
\texttt{{[}a-z{]}+,\ {[}01{]}+} \\
\texttt{x?} & Optional x & \texttt{ab?c,\ {[}+-{]}?} \\
\texttt{(x)} & x, forcing association & \texttt{(aa\textbar{}bb)+} \\
\texttt{x\{n\}} & n occurrences of x & \texttt{{[}0-9{]}\{2\}} \\
\texttt{x\{n,m\}} & n to m occurrences of x & \texttt{a\{2,5\}b*} \\
\texttt{\{xx\}} & Expand xx from the definitions & \texttt{\{digit\}} \\
\end{longtable}

\hypertarget{exercises}{%
\subsection{Exercises}\removelabel{exercises}}

The following exercises start with the above code in file
\texttt{lexer.l} and the final result is a lexical analyser for a
miniature programming language.

\begin{enumerate}
\def\removelabelenumi{\arabic{enumi}.}
\item
  Modify the above code to print the token \emph{value} along with the
  token class. It should, for example, print \texttt{NATURAL(10)} when
  given \texttt{10} as input (and similarly for decimals). Notice that
  \emph{lex} provides an external variable named \texttt{yytext} that
  points to the current string matched by the lexer.
\item
  In programming languages, the names of variables and functions are
  generically referred to as \emph{identifiers}. Modify the code to
  print \texttt{IDENTIFIER(x)} whenever an identifier
  \texttt{\textquotesingle{}x\textquotesingle{}} is found. Identifiers
  consist of non-empty sequences of letters and digits, starting with a
  letter.
\item
  \emph{Keywords} are reserved and cannot be used as identifiers. Modify
  the code to recognize the following tokens: \texttt{integer},
  \texttt{double}, \texttt{if}, \texttt{then} and \texttt{else}. It
  should output \texttt{INTEGER}, \texttt{DOUBLE}, and so on. The key is
  to understand that \emph{lex} always looks for the longest match and,
  in case there is a tie, it chooses the rule that comes first.
\item
  Whitespace is ignored in most languages (Python is a notable
  exception). Modify the code to ignore whitespace characters: spaces,
  tabs and newlines. This can be achieved by matching those characters
  to an empty action \texttt{\{;\}} that simply does nothing.
\item
  Programming languages use punctuation marks with specific meanings.
  Modify the code to recognize the following tokens:
  \texttt{(\ )\ =\ ,\ *\ /\ +\ -}
\item
  A final rule is included to match any other character that could not
  be recognized. This rule must necessarily be the last one. Modify the
  code to show an error message whenever an unrecognized character is
  found. The key is to add a rule for
  \texttt{.\ \{\ printf("error...");\ \}} that will match \emph{any
  single character that has not been matched by other rules}. The error
  message should show the line and column numbers. For this, you will
  need to add variables to the declarations section, which may include C
  code delimited by \texttt{\%\{\ ...\ \%\}}, and update the column
  according to the external variable \texttt{yyleng} that stores the
  length of the token pointed to by \texttt{yytext}.
\end{enumerate}

Finally, test the complete lexical analyser on the following input:

\begin{verbatim}
factorial(integer n) =
    if n then n * factorial(n-1) else 1
    #
\end{verbatim}

The lexer should output the 19 tokens, followed by an error message on
line 3, column 5, because \texttt{\#} is an invalid character preceded
by 4 spaces. The following output is expected:

\begin{verbatim}
IDENTIFIER(factorial)
(
INTEGER
IDENTIFIER(n)
)
=
IF
IDENTIFIER(n)
THEN

...

Line 3, column 5: unrecognized character (#)
\end{verbatim}

\hypertarget{author}{%
\subsection{Author}\removelabel{author}}

Raul Barbosa
(\href{https://apps.uc.pt/mypage/faculty/uc26844}{University of
Coimbra})

\hypertarget{references}{%
\subsection{References}\removelabel{references}}

Niemann, T. (2016) Lex \& Yacc. https://epaperpress.com/lexandyacc

Levine, J. (2009). Flex \& Bison: Text processing tools. O'Reilly Media.

Barbosa, R. (2023). Petit programming language and compiler.\\
https://github.com/rbbarbosa/Petit

Aho, A. V. (2006). Compilers: Principles, techniques and tools, 2nd
edition. Pearson Education.

\newpage
\hypertarget{compilers-tutorial-ii-advanced-lex-features}{%
\section{\texorpdfstring{Compilers tutorial II: Advanced \emph{lex}
features}{Compilers tutorial II: Advanced lex features}}\removelabel{compilers-tutorial-ii-advanced-lex-features}}

\emph{Start conditions} are used to specify different \emph{states} in
which the lexical analyzer can be, based on specific rules and patterns.
Each start condition includes a distinct set of regular expressions that
are active when that condition is triggered. By using start conditions,
\emph{lex} allows for more flexible and modular specification of token
recognition based on the lexical context.

To declare new start conditions, place their names in the declarations
section of the \emph{lex} source file:

\begin{verbatim}
%X STATE1 STATE2
\end{verbatim}

To use a start condition in the rules section, enclose its name in
\texttt{\textless{}\textgreater{}} at the beginning of a rule (the
expression is only matched if the automaton is in that state):

\begin{verbatim}
<STATE1>expression    { printf("found expression in state 1"); }
\end{verbatim}

To move \emph{lex} to a specific start condition, execute
\texttt{BEGIN(condition)} in the action of a rule. To return to the
default initial state of \emph{lex}, execute \texttt{BEGIN(INITIAL)} or
simply \texttt{BEGIN(0)}. Notice that you may remove the parentheses and
simply write \texttt{BEGIN\ condition} and \texttt{BEGIN\ 0}, if you
prefer.

\hypertarget{an-example}{%
\subsection{An example}\removelabel{an-example}}

To exemplify the usage of \emph{lex} states, consider the following
specification:

\begin{verbatim}
%X COMMENT
%%
.                       {;}
"/*"                    { BEGIN(COMMENT); }
<COMMENT>.              { ECHO; }
<COMMENT>\n             { printf(" "); }
<COMMENT>"*/"           { BEGIN(INITIAL); }
%%
extern int yylex();
int main() {
    yylex();
    return 0;
}
int yywrap() {
    return 1;
}
\end{verbatim}

This \emph{lex} specification ignores everything \emph{except} comments.
We could use this to check the spelling of text in code comments.

The first rule discards any character (except newline). The second rule
matches \texttt{/*} and moves the lexical analyser to the
\texttt{COMMENT} state. In that state, any character (except newline) is
printed while newline characters are replaced by spaces. Finally, when
the closing \texttt{*/} is matched, the analyser moves back to the
initial state.

\hypertarget{pre-declared-functions-and-variables}{%
\subsection{Pre-declared functions and
variables}\removelabel{pre-declared-functions-and-variables}}

For reference, the following table summarises the most relevant features
of \emph{lex}.

\begin{longtable}[]{@{}
  >{\raggedright\arraybackslash}p{(\columnwidth - 2\tabcolsep) * \real{0.2899}}
  >{\raggedright\arraybackslash}p{(\columnwidth - 2\tabcolsep) * \real{0.7101}}@{}}
\toprule\noalign{}
\begin{minipage}[b]{\linewidth}\raggedright
Name
\end{minipage} & \begin{minipage}[b]{\linewidth}\raggedright
Description
\end{minipage} \\
\midrule\noalign{}
\endhead
\bottomrule\noalign{}
\endlastfoot
\texttt{int\ yylex(void)} & Call the lexical analyser \\
\texttt{char\ *yytext} & Pointer to the matched token \\
\texttt{yyleng} & Length of the matched token \\
\texttt{yylval} & Semantic value associated with a token \\
\texttt{YY\_USER\_ACTION} & Macro executed before every matched rule's
action \\
\texttt{ECHO} & Print the matched string \\
\texttt{int\ yywrap(void)} & Called on end-of-file, return 1 to stop \\
\texttt{BEGIN\ condition} & Switch to a specific start condition \\
\texttt{INITIAL} & The default initial start condition (same as 0) \\
\texttt{\%X\ condition(s)} & Declare the names of exclusive start
conditions \\
\end{longtable}

All of these features should be familiar to \emph{lex} users. An
advanced feature that can simplify \emph{lex} specifications is the
\texttt{YY\_USER\_ACTION} macro: if we \texttt{\#define} this macro, the
corresponding code will be executed before every single action.
Therefore, it is useful when the same code is repeated in all actions.

\hypertarget{exercises}{%
\subsection{Exercises}\removelabel{exercises}}

The following exercises start with \emph{your} solution to the previous
exercises in file \texttt{lexer.l}. Alternatively, you could also use
the original \texttt{lexer.l} file.

\begin{enumerate}
\def\removelabelenumi{\arabic{enumi}.}
\tightlist
\item
  Many programming languages have block comments delimited by
  \texttt{/*\ ...\ */} that are allowed to span multiple lines. Modify
  the lexical analyser to support block comments. Specifically, it
  should discard all comments while maintaining the line and column
  numbers correctly updated. For simplicity, unterminated comments are
  allowed in our miniature programming language.
\end{enumerate}

Test the lexical analyser on the following input:

\begin{verbatim}
factorial(integer n) =
    if n then n * factorial(n-1) else 1  /* recursive factorial
 */ #
\end{verbatim}

The lexer should output the 19 tokens, followed by an error message on
line 3, column 5, because \texttt{\#} is an invalid character.

\begin{enumerate}
\def\removelabelenumi{\arabic{enumi}.}
\setcounter{enumi}{1}
\item
  Modify the lexical analyser to recognize strings. For example, it
  should print \texttt{STRLIT("hello\textbackslash{}n")} when given
  \texttt{"hello\textbackslash{}n"} as input. Strings are sequences of
  characters (except ``carriage return'', ``newline'' and double
  quotation marks) and/or ``escape sequences'' delimited by double
  quotation marks. Escape sequences \texttt{\textbackslash{}f},
  \texttt{\textbackslash{}n}, \texttt{\textbackslash{}r},
  \texttt{\textbackslash{}t}, \texttt{\textbackslash{}\textbackslash{}}
  and \texttt{\textbackslash{}"} are allowed, while any other escape
  sequences should show an error message like:

  \texttt{Line\ x,\ column\ y:\ invalid\ escape\ sequence\ (\textbackslash{}z)}
\end{enumerate}

\hypertarget{author}{%
\subsection{Author}\removelabel{author}}

Raul Barbosa
(\href{https://apps.uc.pt/mypage/faculty/uc26844}{University of
Coimbra})

\hypertarget{references}{%
\subsection{References}\removelabel{references}}

Levine, J. (2009). Flex \& Bison: Text processing tools. O'Reilly Media.

Niemann, T. (2016) Lex \& Yacc. https://epaperpress.com/lexandyacc

Barbosa, R. (2023). Petit programming language and compiler.\\
https://github.com/rbbarbosa/Petit

Aho, A. V. (2006). Compilers: Principles, techniques and tools, 2nd
edition. Pearson Education.

\newpage
\hypertarget{compilers-tutorial-iii-syntactic-analysis}{%
\section{Compilers tutorial III: Syntactic
analysis}\removelabel{compilers-tutorial-iii-syntactic-analysis}}

\emph{Yacc} is a powerful tool for automating the creation of parsers,
also known as syntactic analysers, mainly used in language processing
and compiler construction. It transforms formal grammars into executable
code, making it invaluable for language analysis and processing. It
primarily supports context-free grammars.

A bit of theory: \emph{Yacc} takes the user-specified grammar rules and
algorithmically constructs the corresponding LR parsing table.
Specifically, it constructs an LALR(1) parser, that is, a Look Ahead
Left-to-right Rightmost derivation parser. The parser takes as input the
sequence of tokens passed by the lexical analyser (\emph{lex}) and moves
according to the parsing table. If it completes the derivation of the
input sequence, reaching an accepting state, then the input is in the
language of the grammar. Otherwise, a syntax error is found.

\hypertarget{using-lex-and-yacc-together}{%
\subsection{\texorpdfstring{Using \emph{lex} and \emph{yacc}
together}{Using lex and yacc together}}\removelabel{using-lex-and-yacc-together}}

\emph{Lex} and \emph{yacc} were designed to work together: \emph{lex}
produces tokens that can be given as input into \emph{yacc}. A
\emph{yacc} source file has three sections separated with the
\texttt{\%\%} delimiter:

\begin{verbatim}
...definitions...
%% 
...rules...
%% 
...subroutines...
\end{verbatim}

The \emph{definitions} section contains C code delimited by
\texttt{\%\{\ ...\ \%\}} and token declarations. The \emph{rules}
section contains the grammar, in a notation similar to BNF (Backus-Naur
form). The \emph{subroutines} section contains any C functions needed.
Integrating \emph{lex} and \emph{yacc} is achieved by placing
\texttt{\%token} declarations in the definitions section of the
\emph{yacc} source file:

\begin{verbatim}
%token NATURAL
\end{verbatim}

This line declares a \texttt{NATURAL} token. The parser generated by
\emph{yacc} is written to a file called \texttt{y.tab.c}, along with an
include file called \texttt{y.tab.h}. The \emph{lex} specification must
include \texttt{y.tab.h} so it can \emph{return} tokens to the parser:

\begin{verbatim}
{digit}+                { yylval = atoi(yytext); return NATURAL; }
\end{verbatim}

To obtain tokens, \emph{yacc} calls the \texttt{yylex()} function, which
returns an integer representing the identified token. The token
\emph{value} is passed to \emph{yacc} in variable \texttt{yylval}. The
type of \texttt{yylval} is \texttt{int} by default. For single-character
tokens there is a convenient shortcut:

\begin{verbatim}
[()=,*/+-]              { return yytext[0]; }
\end{verbatim}

With this rule, \emph{lex} returns the character value for any of those
single-character tokens. It would be functionally equivalent to declare
a \texttt{\%token} for each of them.

\hypertarget{a-simple-calculator-example}{%
\subsection{A simple calculator
example}\removelabel{a-simple-calculator-example}}

To use \emph{yacc}, we write a grammar with fragments of C code enclosed
within curly braces, called actions, attached to the grammar rules.
Those code fragments are executed whenever a rule is used during the
parsing of a sequence of tokens.

A grammar rule has a single nonterminal on the left-hand side of a
production, followed by \texttt{:} and followed by the right-hand side
of the rule, possibly followed by alternative rules using
\texttt{\textbar{}} as a separator. To illustrate this with an example,
consider a small calculator capable of evaluating expressions such as
\texttt{2+3*4} and printing the result. The rules section consists of
the following grammar:

\begin{verbatim}
calculator: expression                  { printf("%d\n", $1); }
          ;

expression: NATURAL                     { $$ = $1; }
          | expression '+' expression   { $$ = $1 + $3; }
          | expression '-' expression   { $$ = $1 - $3; }
          | expression '*' expression   { $$ = $1 * $3; }
          | expression '/' expression   { $$ = $1 / $3; }
          ;
\end{verbatim}

The parser maintains two stacks: a \emph{parse stack} and a \emph{value
stack}. The parse stack represents the current state of the parser,
consisting of terminals and nonterminals. The value stack associates a
value with each element of the parse stack. For instance, when the
parser \emph{shifts} a \texttt{NATURAL} token to the parse stack, the
corresponding \texttt{yylval} is pushed to the value stack.

When the parser \emph{reduces} its stack, by popping the right-hand side
of a production and pushing back the left-hand side nonterminal, the
corresponding action is executed. For example, when the rule
\texttt{expression:\ expression\ \textquotesingle{}+\textquotesingle{}\ expression}
is applied, it will pop the three elements
\texttt{expression\ \textquotesingle{}+\textquotesingle{}\ expression}
and push back \texttt{expression}. The action
\texttt{\{\ \$\$\ =\ \$1\ +\ \$3;\ \}} will be executed.

Notice that the C code of actions can reference positions in the value
stack, by using \texttt{\$1,\ \$2,\ \$3,\ ...,\ \$n} to reference the
values of the right-hand side of the production, that is, the \texttt{n}
values popped from the stack. Moreover, \texttt{\$\$} references the new
value to be placed at the top of the stack. Therefore, the action
\texttt{\{\ \$\$\ =\ \$1\ +\ \$3;\ \}} adds the values of two
expressions and pushes back the resulting sum. This way, the two stacks
remain synchronized.

\hypertarget{generating-and-running-the-parser}{%
\subsection{Generating and running the
parser}\removelabel{generating-and-running-the-parser}}

Having the \emph{yacc} specification in a file named \texttt{calc.y}, we
obtain the C code for the parser by entering:

\begin{verbatim}
$ yacc -dv calc.y
\end{verbatim}

Having the lexical specification in a file named \texttt{calc.l}, we
supply a lexical analyser to read the input and pass the sequence of
tokens to the parser:

\begin{verbatim}
$ lex calc.l
\end{verbatim}

The source code generated by \emph{yacc} is written to \texttt{y.tab.c}
and \texttt{y.tab.h}, while the source code generated by \emph{lex} is
written to \texttt{lex.yy.c}. Compile and run the parser using:

\begin{verbatim}
$ cc y.tab.c lex.yy.c -o calc
$ ./calc
2+3*4              [input]
14                 [output]
\end{verbatim}

There are some important observations about the \emph{lex} and
\emph{yacc}:

\begin{itemize}
\tightlist
\item
  The parser is executed by calling \texttt{yyparse()}, which internally
  calls \texttt{yylex()} to obtain the tokens from the lexical analyser.
\item
  The lexical analyser must \texttt{\#include\ "y.tab.h"} to access the
  \texttt{\%token} declarations specified in the syntactic analyser, so
  it can \emph{return} them.
\item
  The \texttt{yylval} variable is shared between \emph{lex} and
  \emph{yacc} to allow for token \emph{values} to be passed from the
  lexer to the parser.
\item
  When the parser finds a syntax error, it calls the
  \texttt{yyerror(char\ *)} function which we should supply.
\end{itemize}

\hypertarget{grammar-ambiguity}{%
\subsection{Grammar ambiguity}\removelabel{grammar-ambiguity}}

The above grammar is ambiguous. For example, it allows for
\texttt{3-2-1} to be parsed both as \texttt{(3-2)-1} and
\texttt{3-(2-1)}. An ambiguous grammar can't be LR(k) so, more
specifically, it can't be LALR(1). As a result, \emph{yacc} warns of 16
shift/reduce conflicts for the above grammar. It will still generate a
parser which uses \emph{shift} as the default operation.

The command \texttt{yacc\ -dv\ calc.y} includes the \texttt{v} option,
so it produces an extra text file with extension \texttt{.output} that
gives verbose details on the LALR(1) states and all the conflicts. By
inspecting the \texttt{.output} file it is possible to look at
individual states and the operation done for each look-ahead token.

The conflicts are due to the unspecified \emph{associativity} and
\emph{precedence} of operators. Arithmetic operators are generally
\emph{left associative} following the convention of evaluating
expressions in a left-to-right manner. Another convention establishes
that the multiplication \texttt{\textquotesingle{}*\textquotesingle{}}
and division \texttt{\textquotesingle{}/\textquotesingle{}} operators
have higher \emph{precedence} than the addition
\texttt{\textquotesingle{}+\textquotesingle{}} and subtraction
\texttt{\textquotesingle{}-\textquotesingle{}} operators. Imposing these
conventions could be achieved by rewriting the grammar to introduce
terms and factors; more conveniently, \emph{yacc} provides a way to
specify how to solve conflicts without modifying the grammar, which we
examine in the exercises below.

\hypertarget{exercises}{%
\subsection{Exercises}\removelabel{exercises}}

\begin{enumerate}
\def\removelabelenumi{\arabic{enumi}.}
\item
  Modify the specification in \texttt{calc.y} to allow for the usage of
  parentheses to explicitly specify the order of evaluation in
  expressions.
\item
  The grammar is ambiguous and the calculations are often incorrect.
  Test the program with \texttt{4*3-2}, for example. Solve all of the
  shift/reduce conflicts by specifying the precedence and associativity
  of operators.
\end{enumerate}

We can specify the \emph{precedence} and \emph{associativity} of
operators, simultaneously, using the keywords \texttt{\%left},
\texttt{\%right} and \texttt{\%nonassoc} in the definitions section. For
instance,
\texttt{\%left\ \textquotesingle{}+\textquotesingle{}\ \textquotesingle{}-\textquotesingle{}}
states that the operators \texttt{\textquotesingle{}+\textquotesingle{}}
and \texttt{\textquotesingle{}-\textquotesingle{}} are left associative
and have equal precedence. Right associative operators use
\texttt{\%right} and non-associative operators use \texttt{\%nonassoc}.
Consider this example:

\begin{verbatim}
%left LOW
%left '+' '-'
%left HIGH
\end{verbatim}

Operations are listed in order of increasing precedence. Operations
listed on the same line have the same precedence. Therefore, this
example specifies that \texttt{LOW} has a lower precedence than
\texttt{\textquotesingle{}+\textquotesingle{}} and
\texttt{\textquotesingle{}-\textquotesingle{}}, which in turn have a
lower precedence than \texttt{HIGH}.

\begin{enumerate}
\def\removelabelenumi{\arabic{enumi}.}
\setcounter{enumi}{2}
\item
  Modify the grammar to allow for the calculation of multiple
  independent expressions separated by commas. For example, entering
  \texttt{3-2-1,4*3-2,5*5/1*4} should output \texttt{0}, \texttt{10},
  \texttt{100}.
\item
  Modify the grammar to accept if-then-else expressions that behave
  exactly like the ternary operator \texttt{?:} existing in C, Java and
  other programming languages. The syntax is: \texttt{if}
  \emph{expression} \texttt{then} \emph{expression} \texttt{else}
  \emph{expression} (we need the tokens \texttt{IF}, \texttt{THEN} and
  \texttt{ELSE} from the lexical analysis exercises).
\end{enumerate}

Notice that this modification introduces shift/reduce conflicts which
can be solved by using the keyword \texttt{\%prec} associated to the new
grammar rule. Specifically, \texttt{\%prec} appears after the rule,
followed by a token or literal, and specifies that the precedence of the
rule should be the precedence of the token or literal. In this case, the
if-then-else rule should have the lowest precedence.

\begin{enumerate}
\def\removelabelenumi{\arabic{enumi}.}
\setcounter{enumi}{4}
\tightlist
\item
  Modify the code to show the line and column numbers of syntax errors.
\end{enumerate}

\hypertarget{author}{%
\subsection{Author}\removelabel{author}}

Raul Barbosa
(\href{https://apps.uc.pt/mypage/faculty/uc26844}{University of
Coimbra})

\hypertarget{references}{%
\subsection{References}\removelabel{references}}

Niemann, T. (2016) Lex \& Yacc. https://epaperpress.com/lexandyacc

Levine, J. (2009). Flex \& Bison: Text processing tools. O'Reilly Media.

Barbosa, R. (2023). Petit programming language and compiler.\\
https://github.com/rbbarbosa/Petit

Aho, A. V. (2006). Compilers: Principles, techniques and tools, 2nd
edition. Pearson Education.

\newpage
\hypertarget{compilers-tutorial-iv-abstract-syntax}{%
\section{Compilers tutorial IV: Abstract
syntax}\removelabel{compilers-tutorial-iv-abstract-syntax}}

\emph{Yacc} is a powerful tool for generating parsers, by transforming
formal grammars into executable code. In this tutorial we explore how to
construct an abstract syntax tree (AST), which is a kind of parse tree
which discards irrelevant details while fully preserving the meaning of
the original program.

A bit of theory: A syntax-directed translation (SDT) scheme consists of
a grammar with attached program fragments (called \emph{actions}).
Whenever a \emph{production} is used, during syntax analysis, its
\emph{action} is executed. One of the main uses of these schemes is to
build syntax trees. Every time a production is used during bottom-up
parsing, the corresponding action can \emph{create} new nodes and/or
\emph{relate} children nodes to the parent.

\hypertarget{abstract-syntax-trees}{%
\subsection{Abstract syntax trees}\removelabel{abstract-syntax-trees}}

Trees can be represented in a number of ways. The data structures below
specify that a \emph{node} in the AST has a linked list of
\emph{children nodes}:

\begin{verbatim}
struct node {
    enum category category;
    char *token;
    struct node_list *children;
};

struct node_list {
    struct node *node;
    struct node_list *next;
};
\end{verbatim}

Every node has a syntactic \emph{category} denoting a specific
programming construct occurring in the input program, such as a
function, a parameter declaration, or a natural constant:

\begin{verbatim}
enum category {Program, Function, ..., Identifier, Natural, ...};
\end{verbatim}

Every node also includes a pointer to the original \emph{token}. The
token is necessary because a node of category \texttt{Identifier} must
have the name of the function or variable, a node of category
\texttt{Natural} must have the string of digits representing the natural
constant, and so on.

To build an AST we only need two operations: creating a new node and
adding a child node to a parent node. Two functions provide that
functionality:

\begin{verbatim}
struct node *newnode(enum category category, char *token);
void addchild(struct node *parent, struct node *child);
\end{verbatim}

The first function returns a newly allocated node with all its fields
initialized (including an empty list of children). The second function
appends a node to the list of children of the parent node. They are
provided in files
\href{https://github.com/rbbarbosa/Petit/blob/main/tutorial/p4_source/ast.h}{\texttt{ast.h}}
and
\href{https://github.com/rbbarbosa/Petit/blob/main/tutorial/p4_source/ast.c}{\texttt{ast.c}}.

\hypertarget{syntax-directed-translation}{%
\subsection{Syntax-directed
translation}\removelabel{syntax-directed-translation}}

During bottom-up parsing, an action \texttt{\{...\}} is executed when
the corresponding production is used. The combined result of all those
executions is the AST.

Consider the following \emph{yacc} specification (which you will
complete). Notice that it is included in file
\href{https://github.com/rbbarbosa/Petit/blob/main/tutorial/p4_source/petit.y}{\texttt{petit.y}},
that you should carefully analyze.

\begin{verbatim}
program: IDENTIFIER '(' parameters ')' '=' expression
                { $$ = program = newnode(Program, NULL);
                  struct node *function = newnode(Function, NULL);
                  addchild(function, newnode(Identifier, $1));
                  addchild(function, $3);
                  addchild(function, $6);
                  addchild($$, function); }
    ;
parameters: parameter               { /* ... */ }
    | parameters ',' parameter      { /* ... */ }
    ;
parameter: INTEGER IDENTIFIER       { /* ... */ }
    | DOUBLE IDENTIFIER             { /* ... */ }
    ;
arguments: expression               { /* ... */ }
    | arguments ',' expression      { /* ... */ }
    ;
expression: IDENTIFIER              { /* ... */ }
    | NATURAL                       { $$ = newnode(Natural, $1); }
    | DECIMAL                       { /* ... */ }
    | IDENTIFIER '(' arguments ')'  { /* ... */ }
    | IF expression THEN expression ELSE expression  %prec LOW
                                    { /* ... */ }
    | expression '+' expression     { /* ... */ }
    | expression '-' expression     { /* ... */ }
    | expression '*' expression     { /* ... */ }
    | expression '/' expression     { /* ... */ }
    | '(' expression ')'            { $$ = $2; }  
    ;
\end{verbatim}

When the first production is used, the right-hand side contains a
\emph{function} with its \texttt{parameters} and \texttt{expression}.
Parsing will be finishing there. The corresponding action executes 6
statements in the following order: \emph{(1)} the AST's root node is
created (\texttt{program} is declared as a global variable); \emph{(2)}
a new \texttt{Function} node is created; \emph{(3)} a new
\texttt{Identifier} node is created, with the function name, and becomes
a child of the \texttt{Function} node; \emph{(4)} the
\texttt{parameters} node \texttt{\$3} becomes a child of the
\texttt{Function} node; \emph{(5)} the \texttt{expression} node
\texttt{\$6} becomes a child of the \texttt{Function} node; and
\emph{(6)} the new \texttt{Function} node becomes a child of the
\texttt{Program} node.

\hypertarget{token-types-and-union}{%
\subsection{\texorpdfstring{Token types and
\texttt{\%union}}{Token types and \%union}}\removelabel{token-types-and-union}}

As seen in the previous tutorial, the semantic value of a token must be
stored in global variable \texttt{yylval}, which has type \texttt{int}
by default. Token declarations use:

\begin{verbatim}
%token INTEGER DOUBLE IF THEN ELSE
\end{verbatim}

When we need to use multiple data types, the \texttt{\%union}
declaration allows us to specify the distinct types that might be stored
in \texttt{yylval}. In our case:

\begin{verbatim}
%union{
    char *lexeme;
    struct node *node;
}
\end{verbatim}

This \texttt{\%union} declaration modifies the type of \texttt{yylval}
so that it may hold a \emph{lexeme} (\texttt{char\ *}) or a \emph{node}
(\texttt{struct\ node\ *}). In other words, \texttt{yylval} might be a
string or an AST node, and a C union encapsulates the two alternatives.

Then, when we declare a token, the C type is specified as follows:

\begin{verbatim}
%token<lexeme> IDENTIFIER NATURAL DECIMAL
\end{verbatim}

Identifiers, naturals and decimals require their semantic value (the
\texttt{char\ *} to the original string) to be stored. The \emph{lex}
specification should copy the semantic value by executing
\texttt{yylval.lexeme\ =\ strdup(yytext);} before returning any of these
tokens.

Furthermore, syntactic variables (i.e., nonterminals) are specified
using the \texttt{\%type} declaration:

\begin{verbatim}
%type<node> program parameters parameter arguments expression
\end{verbatim}

This way, the type of \texttt{\$\$}, \texttt{\$1}, \texttt{\$2}, etc.,
is correctly handled during parsing.

\hypertarget{exercises}{%
\subsection{Exercises}\removelabel{exercises}}

Begin by carefully examining the file
\href{https://github.com/rbbarbosa/Petit/blob/main/tutorial/p4_source/petit.y}{\texttt{petit.y}}
and the AST construction functions in files
\href{https://github.com/rbbarbosa/Petit/blob/main/tutorial/p4_source/ast.h}{\texttt{ast.h}}
and
\href{https://github.com/rbbarbosa/Petit/blob/main/tutorial/p4_source/ast.c}{\texttt{ast.c}}.

\begin{enumerate}
\def\removelabelenumi{\arabic{enumi}.}
\item
  Complete the actions marked with \texttt{/*\ ...\ */} so that an AST
  is constructed, for each program, using the supplied functions
  \texttt{newnode(...)} and \texttt{addchild(...)}.
\item
  Write a function to recursively traverse the AST and show its content.
  The goal is to call that function immediately after \texttt{yyparse()}
  to check that the AST is correct. Consider the following pseudocode:
\end{enumerate}

\begin{verbatim}
     show(struct node *node, int depth) {
         print(node->category, node->token, depth)
         foreach child in node->children show(child, depth+1)
     }
\end{verbatim}

Taking
\texttt{factorial(integer\ n)\ =\ if\ n\ then\ n\ *\ factorial(n-1)\ else\ 1}
as input, the solution to exercises 1 and 2 should have the following
output:

\begin{verbatim}
  Program
  __Function
  ____Identifier(factorial)
  ____Parameters
  ______Parameter
  ________Integer
  ________Identifier(n)
  ____If
  ______Identifier(n)
  ______Mul
  ________Identifier(n)
  ________Call
  __________Identifier(factorial)
  __________Arguments
  ____________Sub
  ______________Identifier(n)
  ______________Natural(1)
  ______Natural(1)
\end{verbatim}

\begin{enumerate}
\def\removelabelenumi{\arabic{enumi}.}
\setcounter{enumi}{2}
\tightlist
\item
  Modify the grammar to allow for multiple functions, using the
  productions that follow, and implement the necessary action to
  construct the AST.
\end{enumerate}

\begin{verbatim}
     program: IDENTIFIER '(' parameters ')' '=' expression
            | program IDENTIFIER '(' parameters ')' '=' expression
\end{verbatim}

Test your solution with the following example, found in file
\href{https://github.com/rbbarbosa/Petit/blob/main/test/factorial.pt}{\texttt{factorial.pt}}:

\begin{verbatim}
factorial(integer n) = if n then n * factorial(n-1) else 1
main(integer i) = write(factorial(read(0)))
\end{verbatim}

The file
\href{https://github.com/rbbarbosa/Petit/blob/main/tutorial/p4_source/factorial.ast}{\texttt{factorial.ast}}
contains the expected AST for this program.

\hypertarget{author}{%
\subsection{Author}\removelabel{author}}

Raul Barbosa
(\href{https://apps.uc.pt/mypage/faculty/uc26844}{University of
Coimbra})

\hypertarget{references}{%
\subsection{References}\removelabel{references}}

Aho, A. V. (2006). Compilers: Principles, techniques and tools, 2nd
edition. Pearson Education.

Levine, J. (2009). Flex \& Bison: Text processing tools. O'Reilly Media.

Niemann, T. (2016) Lex \& Yacc. https://epaperpress.com/lexandyacc

Barbosa, R. (2023). Petit programming language and compiler.\\
https://github.com/rbbarbosa/Petit

\newpage
\hypertarget{compilers-tutorial-v-semantic-analysis}{%
\section{Compilers tutorial V: Semantic
analysis}\removelabel{compilers-tutorial-v-semantic-analysis}}

Semantic analysis is a crucial phase in the compilation process where
the meaning and validity of a program are examined. It involves a series
of checks and transformations to ensure that the input program adheres
to the rules and constraints of the programming language. Two of the
most important goals are: \emph{scope analysis} and \emph{type
checking}.

A bit of theory: The usual semantic analysis algorithm performs a
depth-first traversal of the AST (abstract syntax tree). It performs
tasks such as annotating the tree with semantic information, type
checking, symbol table construction, scope resolution, and consistency
verification. It ensures that variables and functions are declared
before use, enforces type compatibility, detects and reports errors
related to mismatched types or invalid operations. Semantic analysis
ensures the correctness of the program before proceeding to code
generation.

\hypertarget{scope-analysis}{%
\subsection{Scope analysis}\removelabel{scope-analysis}}

A \emph{symbol table} is constructed by the compiler to store
information about identifiers (e.g., variables, functions) and their
attributes. It tracks the scope, data types, and other properties
associated with each identifier.

\begin{verbatim}
struct symbol_list {
    char *identifier;
    enum type type;
    struct node *node;
    struct symbol_list *next;
};
\end{verbatim}

The \texttt{symbol\_list} structure represents a symbol table as a
linked list where each entry contains: the name of the
\emph{identifier}, its \emph{data type}, and a pointer to the \emph{AST
node} in which it is declared. There are two data types and a special
case:

\begin{verbatim}
enum type {integer_type, double_type, no_type};
\end{verbatim}

To handle a symbol table we only need two operations: \emph{inserting} a
new symbol and \emph{looking up} a symbol by its name. Two functions
provide that functionality:

\begin{verbatim}
insert_symbol(symbol_table, identifier, type, node);
search_symbol(symbol_table, identifier);
\end{verbatim}

The first function inserts a symbol in the table, unless it is already
there. The second function looks up a symbol by its identifier name,
returning \texttt{NULL} if not found. The functions are provided in the
\href{https://github.com/rbbarbosa/Petit/blob/main/tutorial/p5_source/semantics.h}{\texttt{semantics.h}}
and
\href{https://github.com/rbbarbosa/Petit/blob/main/tutorial/p5_source/semantics.c}{\texttt{semantics.c}}
files.

\hypertarget{type-checking}{%
\subsection{Type checking}\removelabel{type-checking}}

The compiler performs \emph{type checking} by traversing the AST, using
the symbol table to check if variables are declared before use, to
enforce type compatibility rules (e.g., addition of numbers but not
strings), to check function calls for argument types, and to validate
language-specific rules regarding data types.

AST nodes must be annotated with their data type. Nodes representing
\emph{expressions} will be annotated with \texttt{integer\_type} or
\texttt{double\_type}, while other nodes will be annotated with
\texttt{no\_type}. Therefore, the AST node data structure must now
include the \texttt{type} field:

\begin{verbatim}
struct node {
    enum category category;
    char *token;
    enum type type;
    struct node_list *children;
};
\end{verbatim}

\hypertarget{the-semantic-analysis-algorithm}{%
\subsection{The semantic analysis
algorithm}\removelabel{the-semantic-analysis-algorithm}}

Semantic analysis is expressed through a recursive traversal of the AST.
Symbol table construction typically occurs during the \emph{descending}
phase, while type calculations and checking usually take place during
the \emph{ascending} phase. However, the order and tasks may vary
depending on the specific design of a compiler.

Function \texttt{check\_program} begins semantic analysis by checking
the AST's root:

\begin{verbatim}
struct symbol_list *symbol_table;

int check_program(struct node *program) {
    symbol_table =
        (struct symbol_list *) malloc(sizeof(struct symbol_list));
    symbol_table->next = NULL;
    struct node_list *child = program->children;
    while((child = child->next) != NULL)
        check_function(child->node);
    return semantic_errors;
}
\end{verbatim}

We declare a global symbol table and initialize it in the
\texttt{check\_program} function. To check a \texttt{Program} node the
compiler recursively checks its children, which are the
\texttt{Function} nodes. It does so for each child by calling
\texttt{check\_function}:

\begin{verbatim}
void check_function(struct node *function) {
    struct node *id = getchild(function, 0);
    if(search_symbol(symbol_table, id->token) == NULL) {
        insert_symbol(symbol_table, id->token, no_type, function);
    } else {
        printf("Identifier %s already declared\n", id->token);
        semantic_errors++;
    }
}
\end{verbatim}

To check a \texttt{Function} node, \texttt{check\_function} must ensure
that its identifier hasn't been declared previously. Notice that we have
a single global symbol table (for now). If the function's identifier
(\texttt{id-\textgreater{}token}) can't be found in the symbol table,
then it is inserted there. Otherwise, a semantic error is printed.

Semantic analysis should next proceed to check the \emph{parameters} of
the function and the \emph{expression} which is forms the body of the
function. To maintain our naming convention, those two checks should be
called \texttt{check\_parameters} and \texttt{check\_expression}. The
exercises below challenge you to continue in this direction by writing
those two checks.

\hypertarget{exercises}{%
\subsection{Exercises}\removelabel{exercises}}

We begin with a simplifying assumption: any identifier may only be used
once, globally. This restriction allows us to use \emph{a single symbol
table for all identifiers}. As a consequence, functions and parameters
cannot share the same name (not even parameters from different
functions). The last exercise contributes to removing this restriction.

\begin{enumerate}
\def\removelabelenumi{\arabic{enumi}.}
\item
  Modify the code to check function \emph{parameters}. Ensure that a
  semantic error is displayed when a parameter's identifier has already
  been declared, using the same error message as for functions.
\item
  Enhance the code to provide the line and column numbers where semantic
  errors are detected. To achieve this, you need to store the line and
  column of lexemes (tokens) in the AST. Two effective approaches for
  this are:

  \begin{itemize}
  \tightlist
  \item
    Modify the lexical analyzer to pass a new \texttt{struct} that
    includes the \texttt{char\ *token} along with its corresponding line
    and column.
  \item
    Enable \texttt{\%locations} in \emph{yacc}, update
    \texttt{yylloc.first\_line} and \texttt{yylloc.first\_column} for
    each token processed by the lexical analyzer, and utilize, for
    instance, \texttt{@2.first\_line} and \texttt{@2.first\_column} to
    reference the location of \texttt{\$2} in a semantic action.
  \end{itemize}
\item
  Revise the code to check function \emph{calls}. Ensure that the
  identifier used is the name of an existing function and that the
  number of arguments in the call matches the number of parameters of
  the function. Otherwise, show appropriate error messages.
\item
  Annotate \emph{expression} nodes with their data type by computing the
  \texttt{type} field during the ascending phase of semantic analysis.
  Begin the process by annotating the leaves falling into the categories
  \texttt{Natural} (\texttt{integer\_type}), \texttt{Decimal}
  (\texttt{double\_type}), and \texttt{Identifier} (retrieved from the
  symbol table). Calculate the type of operations based on a simple
  rule: if the result of the operation involves only one type, it is
  guaranteed to be that specific type; otherwise, it should display an
  error indicating \emph{incompatible types}. When displaying the
  content of the AST, incorporate the type information for each
  \emph{expression} node (e.g., \texttt{Identifier(d)\ -\ double}).
\item
  Revise the code to check for undeclared identifiers. In expressions,
  variables are restricted to the parameters of the respective function.
  To implement this, we shall declare a local scope for each function,
  meaning a dedicated symbol table for each function. One approach is to
  create a new local \texttt{struct\ symbol\_list\ *scope} within
  \texttt{check\_function} and pass this scope down to both
  \texttt{check\_parameters} and \texttt{check\_expression}.
  Subsequently, \texttt{check\_parameters} adds the parameters to the
  local scope, while \texttt{check\_expression} retrieves identifiers
  from the local scope.
\end{enumerate}

\hypertarget{author}{%
\subsection{Author}\removelabel{author}}

Raul Barbosa
(\href{https://apps.uc.pt/mypage/faculty/uc26844}{University of
Coimbra})

\hypertarget{references}{%
\subsection{References}\removelabel{references}}

Aho, A. V. (2006). Compilers: Principles, techniques and tools, 2nd
edition. Pearson Education.

Barbosa, R. (2023). Petit programming language and compiler.\\
https://github.com/rbbarbosa/Petit

Free Software Foundation (2021). GNU Bison Manual, Tracking Locations.
https://www.gnu.org/software/bison/manual/html\_node/Locations.html

\newpage
\hypertarget{compilers-tutorial-vi-code-generation}{%
\section{Compilers tutorial VI: Code
generation}\removelabel{compilers-tutorial-vi-code-generation}}

Once the \emph{analysis} stages are complete, we are ready for the
\emph{synthesis} stage: code generation. The compiler transforms the
abstract syntax tree (AST) into a linear representation, by traversing
the AST and emitting some code at every node. At the core of LLVM there
is the \emph{intermediate representation (IR)} language, which is our
target language for code generation.

\hypertarget{simple-functions-in-llvm-ir}{%
\subsection{Simple functions in LLVM
IR}\removelabel{simple-functions-in-llvm-ir}}

Functions are specified using the \texttt{define} keyword, which has the
following simplified syntax:
\texttt{define\ \textless{}ResultType\textgreater{}\ @\textless{}FunctionName\textgreater{}\ ({[}argument\ list{]})}
followed by the function body. Consider the following example:

\begin{verbatim}
define i32 @avg(i32 %a, i32 %b) {
  %1 = add i32 %a, %b
  %2 = sdiv i32 %1, 2
  ret i32 %2
}

define i32 @main() {
  %1 = call i32 @avg(i32 1, i32 5)
  ret i32 %1
}
\end{verbatim}

First, we define the \texttt{avg} function. The function adds
\texttt{\%a} to \texttt{\%b} and stores the result in temporary register
\texttt{\%1}. Then, it divides \texttt{\%1} by two and stores the result
in temporary \texttt{\%2}, which is the return value. The \texttt{main}
function calls \texttt{avg} with arguments 1 and 5, then storing the
result in \texttt{\%1}, which is the return value. With this code in a
file named \texttt{avg.ll}, we compile and run the program by entering:

\begin{verbatim}
$ llc avg.ll -o avg.s
$ clang avg.s -o avg
$ ./avg
$ echo $?          [print the return code of ./avg]
\end{verbatim}

Alternatively, we can use the LLVM interpreter to achieve the same
result:

\begin{verbatim}
$ lli avg.ll
$ echo $?          [print the return code of lli]
\end{verbatim}

\hypertarget{types}{%
\subsection{Types}\removelabel{types}}

The LLVM IR is a strongly typed assembly language. Primitive types
include integers like \texttt{i32} and \texttt{i64} (32 and 64 bits,
respectively). The \texttt{i1} type is used for booleans. The types
\texttt{float} and \texttt{double} represent floating point numbers (32
and 64 bits, respectively).

Pointers are also allowed (just like in C) so we can use pointers to
primitive types such as \texttt{i32*} and \texttt{double*}, pointers to
pointers such as \texttt{i8**}, pointers to functions such as
\texttt{i32(i32)*}, and pointers to structures.

Arrays are also supported, using the syntax
\texttt{{[}\textless{}N\textgreater{}\ x\ \textless{}Type\textgreater{}{]}},
to represent sequences of elements with a specific type. For example,
\texttt{{[}5\ x\ i8{]}} specifies an array of 5 bytes, and
\texttt{{[}20\ x\ i32{]}} specifies an array of 20 integers.

\hypertarget{operations}{%
\subsection{Operations}\removelabel{operations}}

In the example above, the \texttt{avg} function executes some
operations. The first line adds the two operands and stores the result
in temporary \texttt{\%1}. The addition operation has the syntax
\texttt{\textless{}result\textgreater{}\ =\ add\ \textless{}Type\textgreater{}\ \textless{}op1\textgreater{},\ \textless{}op2\textgreater{}}
where \texttt{\textless{}Type\textgreater{}} is the data type of the
operation (both operands must have the same type). Moreover,
\texttt{\textless{}op1\textgreater{}} and
\texttt{\textless{}op2\textgreater{}} are the two operands for which the
sum is calculated: local variables start with \texttt{\%} while global
variables start with \texttt{@}.

Most instructions return a value that is typically assigned to a
temporary register such as \texttt{\%1}, \texttt{\%2}, or
\texttt{\%tmp}. All LLVM IR instructions are in static single assignment
form (SSA), meaning that each register may only be assigned a value
once. Furthermore, any register must be assigned a value before being
used. To simplify the code, we often use the \texttt{alloca} instruction
to explicitly allocate memory on the stack frame for mutable variables
(shown in the example below).

\hypertarget{calls}{%
\subsection{Calls}\removelabel{calls}}

The \texttt{call} instruction is used for simple function calls, with
explicit arguments, and the \texttt{ret} instruction is used for
returning control flow (and often a value) back to the caller. Examples
can be found above, in the \texttt{main} and \texttt{avg} functions, as
well as below, in the \texttt{factorial} function.

\hypertarget{declaration-of-functions}{%
\subsection{Declaration of functions}\removelabel{declaration-of-functions}}

The \texttt{declare} keyword allows us to declare functions without
specifying their implementation (i.e., without a function body). This
allows us to divide big programs into modules that will be combined
together by the linker. It also allows us to declare external functions
such as those provided by the standard C library (\texttt{printf},
\texttt{puts}, etc.).

Suppose we want an LLVM IR program to call external functions
\href{https://github.com/rbbarbosa/Petit/blob/main/source/io.c}{\texttt{\_read}}
and
\href{https://github.com/rbbarbosa/Petit/blob/main/source/io.c}{\texttt{\_write}}
that are defined in the
\href{https://github.com/rbbarbosa/Petit/blob/main/source/io.c}{\texttt{io.c}}
source file. The LLVM IR program must simply include the following
declarations:

\begin{verbatim}
declare i32 @_read(i32)
declare i32 @_write(i32)
\end{verbatim}

It then becomes possible to call those functions just like any local
function, for example using the instruction
\texttt{\%1\ =\ call\ i32\ @\_write(i32\ 123)} to write number 123 to
the standard output.

\hypertarget{control-flow}{%
\subsection{Control flow}\removelabel{control-flow}}

The \texttt{factorial} example below is the result of compiling the
following program:

\begin{verbatim}
factorial(integer n) = if n then n * factorial(n-1) else 1
\end{verbatim}

This example combines operations, calls, etc., with a new element:
control flow.

\begin{verbatim}
define i32 @factorial(i32 %n) {
  %1 = alloca i32
  %2 = add i32 %n, 0
  %3 = icmp ne i32 %2, 0
  br i1 %3, label %L1then, label %L1else
L1then:
  %4 = add i32 %n, 0
  %5 = add i32 %n, 0
  %6 = add i32 1, 0
  %7 = sub i32 %5, %6
  %8 = call i32 @factorial(i32 %7)
  %9 = mul i32 %4, %8
  store i32 %9, i32* %1
  br label %L1end
L1else:
  %10 = add i32 1, 0
  store i32 %10, i32* %1
  br label %L1end
L1end:
  %11 = load i32, i32* %1
  ret i32 %11
}
\end{verbatim}

The \texttt{factorial} function is divided into 4 blocks of
instructions, delimited by the labels. The \texttt{br} instructions are
used to transfer control flow to a different block within the same
function.

The first \texttt{br} instruction is a \emph{conditional branch}. It
takes a single \texttt{i1} value (the boolean condition) and two labels.
If the \texttt{i1} value is true, control flows to the first label;
otherwise, control flows to the second label. The \texttt{i1} value
(temporary \texttt{\%3}) is computed by the \texttt{icmp} instruction,
which is the integer comparison instruction. Therefore, if \texttt{n} is
not equal to zero (\texttt{icmp\ ne\ i32\ \%2,\ 0} where
\texttt{\%2\ =\ n}) the program branches to label \texttt{L1then},
otherwise it branches to label \texttt{L1else}.

The other two \texttt{br} instructions are \emph{unconditional
branches}, taking a single label as target. Such branches are always
taken, that is, control flow is unconditionally transferred to the
target label.

Finally, it should be noted that the \texttt{factorial} function returns
the value stored at \texttt{i32*\ \%1} which is a pointer to an integer
stored in the stack frame. That pointer is allocated through the first
\texttt{alloca} instruction.

\hypertarget{the-recursive-code-generation-algorithm}{%
\subsection{The recursive code generation
algorithm}\removelabel{the-recursive-code-generation-algorithm}}

The input program is fully represented by an AST annotated with the
relevant attributes. Code generation is expressed through a recursive
traversal of the AST. Function \texttt{codegen\_program} (in file
\href{https://github.com/rbbarbosa/Petit/blob/main/tutorial/p6_source/codegen.c}{\texttt{codegen.c}})
generates code for the root \texttt{Program} node by generating code for
each child \texttt{Function} node and then emitting code for the
\texttt{main} entry point:

\begin{verbatim}
void codegen_program(struct node *program) {
    struct node_list *function = program->children;
    while((function = function->next) != NULL)
        codegen_function(function->node);

    struct symbol_list *entry = search_symbol(symbol_table, "main");
    if(entry != NULL && entry->node->category == Function)
        printf("define i32 @main() {\n"
               "  %%1 = call i32 @_main(i32 0)\n"
               "  ret i32 %%1\n"
               "}\n");
}
\end{verbatim}

The \texttt{while} loop iterates through the children of the
\texttt{Program} node, which are the \texttt{Function} nodes, calling
\texttt{codegen\_function} to generate code for each of them.

\begin{verbatim}
void codegen_function(struct node *function) {
    temporary = 1;
    printf("define i32 @_%s(", getchild(function, 0)->token);
    codegen_parameters(getchild(function, 1));
    printf(") {\n");
    codegen_expression(getchild(function, 2));
    printf("  ret i32 %%%d\n", temporary-1);
    printf("}\n\n");
}
\end{verbatim}

The code above initialises the counter of temporary registers
(\texttt{\%1}, \texttt{\%2}, etc.). Then, it emits code to
\texttt{define} the function, calls \texttt{codegen\_parameters} to
generate the list of parameters and calls \texttt{codegen\_expression}
to evaluate the expression. Small fragments of code are also emitted
(the \texttt{ret} instruction, brackets, etc.).

\begin{verbatim}
void codegen_parameters(struct node *parameters) {
    struct node *parameter;
    int curr = 0;
    while((parameter = getchild(parameters, curr++)) != NULL) {
        if(curr > 1)
            printf(", ");
        printf("i32 %%%s", getchild(parameter, 1)->token);
    }
}
\end{verbatim}

Function \texttt{codegen\_parameters} (above) is quite simple, as it
only emits a sequence of identifier names separated by commas. Function
\texttt{codegen\_expression} (below) generates code to evaluate
expressions:

\begin{verbatim}
int codegen_expression(struct node *expression) {
    int tmp = -1;
    switch(expression->category) {
        case Natural:
            tmp = codegen_natural(expression);
            break;
        default:
            break;
    }
    return tmp;
}
\end{verbatim}

Function \texttt{codegen\_expression} is incomplete, because it only
supports expressions of category \texttt{Natural}, which is the simplest
case. The exercises challenge you to continue the code generation for
the other cases (additions, calls, etc.).

Generating code to evaluate an expression of category \texttt{Natural}
consists of loading the value of the natural number (given by the
lexical token) into a temporary register. That code is generated by the
function \texttt{codegen\_natural} below:

\begin{verbatim}
int codegen_natural(struct node *natural) {
    printf("  %%%d = add i32 %s, 0\n", temporary, natural->token);
    return temporary++;
}
\end{verbatim}

Notice that the function \texttt{codegen\_natural} emits the code to
load the value of the natural number into \emph{the next} temporary
register (using the \texttt{temporary} counter). Then, the number of the
temporary register is returned and incremented.

\hypertarget{exercises}{%
\subsection{Exercises}\removelabel{exercises}}

In this tutorial we only consider the \texttt{integer} type and
completely forget about \texttt{double} values (for now). Hence, all
instructions operate on \texttt{i32} values exclusively.

\begin{enumerate}
\def\removelabelenumi{\arabic{enumi}.}
\tightlist
\item
  Take the above example in file
  \href{https://github.com/rbbarbosa/Petit/blob/main/tutorial/p6_source/factorial.ll}{\texttt{factorial.ll}}
  as a starting point. Manually compose the LLVM IR for the
  \texttt{main} function to read an integer value, calculate its
  factorial, and write the result. The read/write functions are
  available in
  \href{https://github.com/rbbarbosa/Petit/blob/main/source/io.c}{\texttt{io.c}}
  and the commands that follow should correctly output the result.
  Notice that we must call \texttt{read(0)} to read a new value.
\end{enumerate}

\begin{verbatim}
     $ llc factorial.ll -o factorial.s
     $ clang factorial.s io.c -o factorial
     $ ./factorial
     12                 [input]
     479001600          [output]
\end{verbatim}

\begin{enumerate}
\def\removelabelenumi{\arabic{enumi}.}
\setcounter{enumi}{1}
\item
  The code generator already compiles functions like
  \texttt{func(integer\ i)\ =\ 10} where the expression consists of a
  single \texttt{Natural} value. Modify the code generator to support
  the \texttt{Identifier} case of expressions. It should then be able to
  compile the program \texttt{identity(integer\ n)\ =\ n} by generating
  code that reads a variable into a temporary register.
\item
  Implement code generation for the multiplication operation. The code
  generation process for this operation can be outlined with the
  following pseudocode:

  \begin{itemize}
  \tightlist
  \item
    \texttt{t1\ =\ codegen\_expression(left\ child)}
  \item
    \texttt{t2\ =\ codegen\_expression(right\ child)}
  \item
    \texttt{new\_temporary\ =} result of multiplying \texttt{t1\ *\ t2}
  \item
    return \texttt{new\_temporary} and post-increment the temporary
    counter
  \end{itemize}
\item
  Implement code generation for the other three arithmetic operations
  (addition, subtraction and division). At this point it should be able
  to compile \texttt{mod(integer\ a,\ integer\ b)\ =\ a-a/b*b} and
  similar functions.
\item
  Implement code generation for function calls, ensuring the generation
  of calls with appropriate arguments. Please be aware of the convention
  to prefix function names with an underscore.
\item
  Implement code generation for if-then-else expressions, with the same
  exact semantics as the the ternary operator \texttt{?:} existing in C
  and Java.
\end{enumerate}

Finally, test your solution with the following Petit program:

\begin{verbatim}
factorial(integer n) = if n then n * factorial(n-1) else 1
main(integer i) = write(factorial(read(0)))
\end{verbatim}

By following the link in the references below, you can find other Petit
programs such as
\href{https://github.com/rbbarbosa/Petit/blob/main/test/primes.pt}{\texttt{primes.pt}}
to test your solution with more advanced programs.

\hypertarget{author}{%
\subsection{Author}\removelabel{author}}

Raul Barbosa
(\href{https://apps.uc.pt/mypage/faculty/uc26844}{University of
Coimbra})

\hypertarget{references}{%
\subsection{References}\removelabel{references}}

M. Rodler, M. Egevig (2023). Mapping High Level Constructs to LLVM IR.\\
https://mapping-high-level-constructs-to-llvm-ir.readthedocs.io

Aho, A. V. (2006). Compilers: Principles, techniques and tools, 2nd
edition. Pearson Education.

Barbosa, R. (2023). Petit programming language and compiler.\\
https://github.com/rbbarbosa/Petit

LLVM Project (2023). LLVM Language Reference Manual.\\
https://llvm.org/docs/LangRef.html

\end{document}